\documentclass{article}

\usepackage{arxiv}

\usepackage[utf8]{inputenc} 
\usepackage[T1]{fontenc}    
\usepackage{hyperref}       
\usepackage{url}            
\usepackage{booktabs}       
\usepackage{amsfonts}       
\usepackage{nicefrac}       
\usepackage{microtype}      
\usepackage{lipsum}		
\usepackage{graphicx}
\usepackage{natbib}
\usepackage{doi}

\usepackage{amsmath}

\title{Hit ratio: An Evaluation Metric for Hashtag Recommendation}

\author{
 Areej Alsini \\
  The University of Western Australia\\
  \texttt{areej.alsini@research.uwa.edu.au} \\
   \And
 Du Q. Huynh \\
 The University of Western Australia\\
  \texttt{du.huynh@uwa.edu.au} \\
  \And
 Amitava Datta \\
 The University of Western Australia\\
  \texttt{amitava.datta@uwa.edu.au} \\
 
}

\renewcommand{\shorttitle}{\textit{arXiv} Template}

\hypersetup{
pdftitle={A template for the arxiv style},
pdfsubject={q-bio.NC, q-bio.QM},
pdfauthor={David S.~Hippocampus, Elias D.~Striatum},
pdfkeywords={First keyword, Second keyword, More},
}

\begin{document}
\maketitle

\begin{abstract}
Hashtag recommendation is a crucial task, especially with an increase of interest in using social media platforms such as Twitter in the last decade. Hashtag recommendation systems automatically suggest hashtags to a user while writing a tweet. Most of the research in the area of hashtag recommendation have used classical metrics such as \textit{hit rate}, \textit{precision}, \textit{recall}, and F1-score to measure the accuracy of hashtag recommendation systems. These metrics are based on the exact match of the recommended hashtags with their corresponding ground truth. However, it is not clear how adequate these metrics to evaluate hashtag recommendation. The research question that we are interested in seeking an answer is: are these metrics adequate for evaluating hashtag recommendation systems when the numbers of ground truth hashtags in tweets are highly variable? In this paper, we propose a new metric which we call \textit{hit ratio} for hashtag recommendation. Extensive evaluation through hypothetical examples and real-world application across a range of hashtag recommendation models indicate that the hit ratio is a useful metric. A comparison of hit ratio with the classical evaluation metrics reveals their limitations.
\end{abstract}


\section{Introduction}
Hashtag recommendation has gained considerable interest due to the vast amount of hashtags generated over time within social media platforms such as Twitter, Facebook, and Instagram.  In the past decade, researchers have looked into Twitter hashtag recommendation task as a supervised learning classification problem \cite{Mazzia-2011,Weston-2014,Dovgopol-15,Gong-2015-hashtag,LiLiu-2016,LiXu-2016}, or unsupervised learning through semantic similarity of tweets~\cite{Kywe-2012,Wang-2014,Sedhai-2014,Song-2015,Li-2016,Zhao-2016,Kowald-2017,Alsini-2017,Benlhachemi-2018,FeiFei-2018,Alsini-2020} to recommend the top-$k$ hashtags to a user while writing a tweet.  %
To test a hashtag recommendation model on unseen data, researchers consider the list of hashtags used by a user in his/her original tweet as the ground truth hashtags. Each of the hashtags in the ground truth represents a label. As these hashtags are user-generated, the number of labels in the ground truth varies from one tweet to another, i.e., the number of the ground truth labels can be larger or smaller than the $k$ value in the top-$k$ recommended hashtags for different tweets. Evaluation in this setting is known to be an issue of debate, as the number of labels for each test example is variable, and has been a topic of study in the extreme multi-class multi-label classification literature.

Classical evaluation metrics such as \textit{hit rate}, \textit{precision}, \textit{recall}, and F1-score are used to evaluate the performance of hashtag recommendation models~\cite{Mazzia-2011,Weston-2014,Dovgopol-15,Gong-2015-hashtag,LiLiu-2016,LiXu-2016,Kywe-2012,Wang-2014,Sedhai-2014,Song-2015,Li-2016,Zhao-2016,Kowald-2017,Alsini-2017,Benlhachemi-2018,FeiFei-2018,Alsini-2020}. The generic notion of these evaluation metrics is to determine the ratio of the correct recommendations of a given test tweet. The recommendations can be \textit{fully correct}, \textit{partially correct}, or \textit{fully incorrect}~\cite{Sorower-2010}. A recommendation is said to be \textit{fully correct} when the top-$k$ recommended hashtags completely match the hashtags in the ground truth. A \textit{fully incorrect} recommendation refers to the case where no correct matches are identified. \textit{Partially correct} is when one or more hashtags (less than $k$) are identified as correct. However, each of these evaluation metrics measures different quantities. Precision and recall calculate the ratio of the correct hashtags to the top-$k$ recommendations and the number of hashtags in the ground truth, respectively. F1-score is the harmonic average between precision and recall. Hit rate only considers a \textit{hit} if there is at least a single correct hashtag between the recommended and the ground truth hashtags.

The variable number of labels in every test example affects the performance scores calculated using the classical evaluation metrics, especially precision, recall, and F1-score. Thus, there is a need for a new evaluation metric that is more suitable for hashtag recommendations. In this paper, we propose the \textit{hit ratio} metric that calculates the ratio of the hits taking into consideration the variable number of hashtags in the recommended and the ground truth hashtags. We compare the results of hashtag recommendation using the hit ratio with precision, recall, F1-score, and hit rate. Section 2 briefly reviews the basics of evaluation measures, explains our proposed hit ratio metric for the evaluation of hashtag recommendation models, and demonstrates hypothetical examples. Section 3 presents the experiments and compares the hit ratio metric versus classical measures. Conclusions and future works are described in Section 4.

\section{Our Proposed Method}
\subsection{A brief Review of the Basics}
Classical evaluation metrics such as precision, recall, and F1-score are useful metrics for evaluating the accuracy of various applications. In binary classification, \textit{precision} measures the proportion of true-positives in the positive predictions returned by an algorithm whereas \textit{recall} measures the proportion of ground truth positives that are correctly identified by an algorithm. That is, precision is calculated as $TP/(TP + FP)$ and recall is calculated as $TP/(TP + FN)$, where $TP$ is the number of true-positives, $FP$ is the number of false-positives, and $FN$ is the number of false-negative cases. These two measures are critical when the costs of false-positive and false-negative predictions are very high. $FP$ represents a case of a diabetic patient who is diagnosed as normal, and $FN$ represents a case of a normal person who is diagnosed as a diabetic. Then, both cases are treated accordingly. The multi-class evaluation is an extension of the binary class evaluation. It assesses the performance of the predictions for each class separately and averages the overall performance classes. Classical evaluation metrics also can be adopted in multi-class multi-label evaluation, primarily when the number of classes is determined in advance. This method of evaluation is known as \textit{label-based} evaluation~\cite{Sorower-2010}\cite{Forero-2015}.

In multi-label multi-class classification, each label can be considered as an attribute (e.g., "type of fruit"; "colour of fruit") and the multiple classes under each label can be considered as the different values the attribute can have (e.g., classes "apple", "orange", and "banana" for the label "type of fruit"; classes "red", "green", "orange", and "yellow" for the label "colour of fruit"). Although the number of ground truth hashtags varies from one tweet to another tweet, the number of ground truth hashtags should not be interpreted as the number of labels in label-based evaluation, as they are not different attributes of hashtags. Instead, the evaluation of the performance of hashtag recommendation systems is more similar to \textit{example-based} evaluation~\cite{Sorower-2010}\cite{Forero-2015}, where the precision and recall metrics have slightly different meanings. For a single test tweet in hashtag recommendation, precision measures the ratio of the correctly recommended hashtags and recall measures the ratio of the correct hashtags detected from the ground truth. Let $R = \{h_1, h_2, \cdots, h_{n_R}\}$ be the list of the top-$k$ recommended hashtags, where $n_G \leq k$, as explained later. Let $G= \{h'_1, h'_2, \cdots, h'_{n_G}\}$ be the set of ground truth hashtags for a particular tweet. Then, ${m=|G \cap R|}$ is the total number of correctly matching hashtags between these two sets. Thus, in terms of hashtag recommendation, the precision metric is defined as:


\begin{equation}\label{eq:P}
\text{precision} = \frac{m}{k},
\end{equation}


\noindent
and the recall metric is defined as:

\begin{equation}\label{eq:R}
\text{recall} = \frac{m}{n_G}.
\end{equation}

Ideally, the cardinality of the set $R$, i.e., the value $n_R$, should be equal to $k$ for the top-$k$ recommendation. However, in the comparison of hashtags, there may not be sufficient recommended hashtags which match well with the ground truth hashtags, e.g., their cosine similarity scores might be below an acceptable threshold value. So, they are excluded as a recommendation~\cite{Alsini-2020}. In such a case, we have $n_R < k$.

For the hit rate metric, a hit is registered for a single test tweet when there is at least a single common hashtag between the recommended and the ground truth hashtags. For example, if the top-1 recommended hashtag for a given tweet is \textit{\#entrepreneur} and the ground truth hashtag is \textit{\#entrepreneur}, this is scored as 1. This metric, however, does not reflect the quality of the hit. For instance, for top-$k$ recommendation with $k = 5$, say, if all the five recommended hashtags match perfectly with the five ground truth hashtags, then the hit rate is 1. If only one of the five recommended hashtags matches one of the ground truth hashtags, the hit rate is also 1. The quality of the former recommendation is better than the latter. Unfortunately, using the traditional hit rate metric, we cannot distinguish these two cases when $k > 1$.

To calculate the performance of the hashtag recommendation model, the average precision, average recall, and average hit rate are calculated as the number of correct recommendations over the number of test tweets.


\subsection{The Hit Ratio Metric}
Knowing that the values of $n_R$ and $n_G$ vary in different tweets, we consider using a ratio of matching hashtags over the minimum value between them to measure the hit ratio. We define the hit ratio of a single test tweet as: 

\begin{equation}\label{eq:Hit_Ratio}
\text{hit ratio}= \frac{m}{\text{min}(n_R,n_G)}.
\end{equation}
This metric is bounded between [0,1] where 0 means there is no match (fully incorrect) and 1 means all possible matches are predicted (fully correct), and the values in between are the ratio of the match (partially correct). To calculate the overall evaluation score of a model, we average the hit ratio over all the test tweets.

\begin{table*}[h!]
\centering
\resizebox{\textwidth}{!}{%
\begin{tabular}{c|c|c|c|c|c|c|c|c|c|c|c|c|c|c|c|c|c|c|}
\cline{2-19}
                                 & \multicolumn{6}{c|}{\textbf{Example 1}}                                       & \multicolumn{6}{c|}{\textbf{Example 2}}                                       & \multicolumn{6}{c|}{\textbf{Example 3}}                                       \\ \hline
\multicolumn{1}{|c|}{\textbf{$n_G$}} & \textbf{$m$} & \textbf{hit rate} & \textbf{P} & \textbf{R} & \textbf{F1} & \textbf{hit ratio} & \textbf{$m$} & \textbf{hit rate} & \textbf{P} & \textbf{R} & \textbf{F1} & \textbf{hit ratio} & \textbf{$m$} & \textbf{hit rate} & \textbf{P} & \textbf{R} & \textbf{F1} & \textbf{hit ratio} \\ \hline
\multicolumn{1}{|c|}        {1}          & 1          & 1.00              & 0.33       & 1.00     &     0.50         & 1.00                  & 1          & 1.00              & 0.33       & 1.00      &       0.50          & 1.00               & 1          & 1.00              & 0.33       & 1.00     &   0.50          & 1.00               \\ \hline
\multicolumn{1}{|c|}        {2}          & 1          & 1.00              & 0.33       & 0.50     &     0.40         & 0.50                  & 2          & 1.00              & 0.67       & 1.00      &       0.80          & 1.00               & 2          & 1.00              & 0.67       & 1.00     &   0.80          & 1.00               \\ \hline
\multicolumn{1}{|c|}        {3}          & 1          & 1.00              & 0.33       & 0.33     &     0.33         & 0.33                  & 2          & 1.00              & 0.67       & 0.67      &       0.67          & 0.67               & 3          & 1.00              & 1.00       & 1.00     &    1.00         & 1.00               \\ \hline
\multicolumn{1}{|c|}        {4}          & 1          & 1.00              & 0.33       & 0.25     &     0.28         & 0.33                  & 2          & 1.00              & 0.67       & 0.50      &       0.57          & 0.67               & 3          & 1.00              & 1.00       & 0.75     &    0.86        & 1.00               \\ \hline
\multicolumn{1}{|c|}        {5}          & 1          & 1.00              & 0.33       & 0.20      &     0.25         & 0.33                  & 2          & 1.00              & 0.67       & 0.40      &       0.50          & 0.67               & 3          & 1.00              & 1.00       & 0.60     &    0.75        & 1.00               \\ \hline

\end{tabular}%
}
\caption{Examples showing the values of the hit rate, precision, recall, F1-score, and hit ratio when $n_R$ is fixed to 3, $n_G$ is the number of hashtags in the ground truth, and $m$ is the number of matching hashtags.}
\label{tab:example1}
\end{table*}

\begin{table*}[h!]
\centering
\resizebox{\textwidth}{!}{%
\begin{tabular}{c|c|c|c|c|c|c|c|c|c|c|c|c|c|c|c|c|c|c|}
\cline{2-19}
\multicolumn{1}{l|}{}            & \multicolumn{6}{c|}{\textbf{Example 4}}                                       & \multicolumn{6}{c|}{\textbf{Example 5}}                                       & \multicolumn{6}{c|}{\textbf{Example 6}}                                       \\ \hline
\multicolumn{1}{|c|}{\textbf{$n_R$}} & \textbf{$m$} & \textbf{hit rate} & \textbf{P} & \textbf{R} & \textbf{F1} & \textbf{hit ratio} & \textbf{$m$} & \textbf{hit rate} & \textbf{P} & \textbf{R} & \textbf{F1} & \textbf{hit ratio} & \textbf{$m$} & \textbf{hit rate} & \textbf{P} & \textbf{R} & \textbf{F1} & \textbf{hit ratio} \\ \hline
\multicolumn{1}{|c|}    {1}               & 1          & 1.00           & 1.00          & 0.33    &     0.50          & 1.00                  & 1                  & 1.00    & 1.00          & 0.33   &      0.50        & 1.00                  & 1          & 1.00                 & 1.00          & 0.33  &    0.50    & 1.00                  \\ \hline
\multicolumn{1}{|c|}    {2}               & 1          & 1.00            & 0.50          & 0.33    &     0.40          & 0.50                   & 2                  & 1.00    & 1.00          & 0.67   &      0.80        & 1.00                  & 2          & 1.00                 & 1.00          & 0.67  &    0.80    & 1.00                  \\ \hline
\multicolumn{1}{|c|}    {3}               & 1          & 1.00           & 0.33          & 0.33    &     0.33          & 0.33                  & 2                  & 1.00    & 0.67          & 0.67   &      0.67        & 0.67                  & 3          & 1.00                 & 1.00          & 1.00  &    1.00    & 1.00                  \\ \hline
\multicolumn{1}{|c|}    {4}               & 1          & 1.00           & 0.25          & 0.33    &     0.28          & 0.33                  & 2                  & 1.00    & 0.50          & 0.67   &      0.57        & 0.67                  & 3          & 1.00                 & 0.75          & 1.00  &    0.86    & 1.00                  \\ \hline
\multicolumn{1}{|c|}    {5}               & 1          & 1.00           & 0.20          & 0.33    &     0.25          & 0.33                  & 2                  & 1.00    & 0.40          & 0.67   &      0.50        & 0.67                  & 3          & 1.00                 & 0.60          & 1.00  &    0.75   & 1.00                  \\ \hline
\end{tabular}%
} 
\caption{Examples that show the values of the hit rate, precision, recall, F1-score, and hit ratio when $n_R$ ranges between 1 to 5, the number of hashtags $n_G$ is fixed to 3, and $m$ is the number of matching hashtags.}
\label{tab:example2}
\end{table*}

To examine the effectiveness of the hit ratio metric and the limitations of the classical evaluation metrics, we explain the computation through some hypothetical examples. In Table~\ref{tab:example1}, we fix the value of $n_R$ to 3 and vary the values of $n_G$ to range from 1 to 5. In Table~\ref{tab:example2}, we fix the value of $n_G$ and vary $n_R$ instead. We notice the following:

\begin{itemize}
    \item Precision, recall, F1-score, and hit ratio achieve full scores when there is a perfect match between the $n_R$ recommended and $n_G$ ground truth hashtags. This is very clear in Example 3 and 6 where the number of matching hashtags $m$ equals 3, the number of hashtags in the ground truth $n_G$ equals 3, and the number of recommended hashtags $n_R$ equals 3.
    \item Hit rate scores as 1 in all the examples since there is at least one possible match in each of the cases.
    \item The precision value decreases with the increase in the value of the $n_R$ even if all possible matches are obtained. This is shown in Examples 4-6.
    \item The recall value decreases as $n_G$ increases even if all possible matches are achieved. This is noticed in Examples 1-3.
    \item The F1-score varies even though the number of matches is the same. This can be seen in Examples 3 and 6 where $m=3$.  
    \item Hit ratio reaches its maximum value 1 if all possible matches are achieved ($m=\text{min}(n_R,n_G)$) shown in Examples 3 and 6. 
    \item The hit ratio decreases only when the number of matches decreases (i.e. when $m<\text{min}(n_R,n_G)$), as shown in Examples 1, 2, 4, and 5.
\end{itemize}

\section{Experiments}
For the hit ratio metric, there are no prior issues that need to be solved or extra values that need to be calculated. However, to evaluate the effectiveness of a hashtag recommendation model, we need the top-$k$ hashtag recommendations with their corresponding ground truth hashtags. Our aim in this section is to compare the scores calculated using the hit ratio versus the classical evaluation measures in hashtag recommendation and not to compare between the hashtag recommendation models. For this reason, we briefly describe the datasets, pre-processing steps, word embeddings training and hashtag recommendation models that we have adopted in our experiments. 

\subsection{Datasets and Pre-processing}
In our experiments, we use the~\textit{Dataset-UDI-TwitterCrawl-Aug2012}~\cite{UDIdataset-12}. We are only interested in the personal timeline of users who have used at least three hashtags in their historical tweets. As for the pre-processing step of the \textit{hashtaged tweets} (tweets composing hashtags), all tweets are transformed into lower case. We remove stop words, punctuation except for the \# sign, URLs, repetitive words, retweets, and mentions. Hashtagged tweets with less than three words after pre-processing are also removed. Table~\ref{tab:statistics} shows the statistics of the hashtags in the hashtagged tweets. Then, we divide each of the detected communities into a repository and testing sets. Each test set includes the most recent 10\% of the community's hashtagged tweets. 

\begin{table*}[h]
\centering
\resizebox{0.5\textwidth}{!}{%
\begin{tabular}{|l|r|}
\hline
No. of hashtagged tweet after pre-processing  & 6,762,211 \\ \hline
Max. number of hashtags per tweet & 27        \\ \hline
Min. number of hashtags per tweet & 1         \\ \hline
Avg. number of hashtags per tweet & 2         \\ \hline
\end{tabular}%
}
\caption{Statistics of hashtags in the hashtagged tweets.}
\label{tab:statistics}
\end{table*}

\subsection{Training the Word Embeddings Models}
\label{we_training}
In this paper, we have trained two word embeddings models to represent tweets and hashtags: Word2Vec~\cite{Mikolov-2013} and FastText~\cite{athiwaratkun-2018}. We use the same hyperparameters for both models as follows: we set the dimension of words to 300 and the \textit{window size} to 5.

\subsection{Hashtag Recommendation Models}
We have chosen the Community-Based Hashtag Recommendation model proposed in~\cite{Alsini-2020} for its efficiency in terms of processing time. The community-based hashtag recommendation has been investigated using various settings. In this paper, we choose the best performing model, which has its communities detected using the Cliques Percolation Method (CPM) from the network generated using the similarity of hashtag usage between the users. However, we only alter the tweet vectorization method to create three models. For the top-$k$ hashtags to be recommended for a given tweet, similar tweets need to be retrieved. The threshold value of the tweet similarity is set to 0.5. Hashtags are extracted from the set of similar tweets, scored and ranked for the recommendation.

\noindent
\textbf{Model A:} This model uses the TF-IDF method to vectorize its tweets and the \textit{tweet hashtag relevance} as the ranking method of hashtags~\cite{Alsini-2020}.

\noindent
\textbf{Model B:} This model uses the Mean of Words Embeddings MOWE of words trained using the Word2Vec model (Section~\ref{we_training}) to vectorize its tweets and the \textit{hashtag popularity} as the ranking method of hashtags~\cite{Alsini-2020}.

\noindent
\textbf{Model C:} This model uses the Mean of Words Embeddings MOWE of words trained using the FastText model (Section~\ref{we_training}) to vectorize its tweets and the \textit{hashtag popularity} as the ranking method of hashtags.

Then, we compare the average measures of the hit rate, precision, recall, F1-score, and hit ratio of the three models.

\subsection{Results and Discussions}

\begin{table*}[t!]
\centering
\resizebox{\textwidth}{!}{%
\begin{tabular}{|c|c|c|c|c|c|c|c|c|c|c|c|c|c|c|c|}
\hline
\textbf{Recommendation}                       & \multicolumn{5}{c|}{\textbf{Model A}}                            & \multicolumn{5}{c|}{\textbf{Model B}}                            & \multicolumn{5}{c|}{\textbf{Model C}}                            \\ \hline
\textbf{top-k}                           & \textbf{hit rate} & \textbf{P} & \textbf{R} & \textbf{F1}    & \textbf{hit ratio} & \textbf{hit rate} & \textbf{P} & \textbf{R} & \textbf{F1} & \textbf{hit ratio} & \textbf{hit rate} & \textbf{P} & \textbf{R} & \textbf{F1} & \textbf{hit ratio} \\ \hline
top-1                                    & 0.21              & 0.21       & 0.19       &     0.2           & 0.20               & 0.13              & 0.13       & 0.10     &      0.11         & 0.13               & 0.13              & 0.13       & 0.10       &      0.11       & 0.12               \\ \hline
top-5                                    & 0.21              & 0.05       & 0.19       &     0.08          & 0.20               & 0.27              & 0.06       & 0.24     &      0.10         & 0.24               & 0.27              & 0.06       & 0.24       &      0.10       & 0.24               \\ \hline
top-10                                   & 0.21              & 0.03       & 0.19       &     0.05          & 0.20               & 0.34              & 0.04       & 0.30     &      0.07         & 0.30               & 0.34              & 0.04       & 0.30       &      0.07       & 0.30               \\ \hline

\end{tabular}%
}
\caption{A comparison between measures of hashtag recommendation.}
\label{tab:practice_results}
\end{table*}

Table~\ref{tab:practice_results} shows a comparison between measures of hashtag recommendation. We notice the following:
\begin{itemize}
    \item The three models A, B, and C achieve recall scores closer to the hit ratio.
    \item For the three models, precision has the lowest score than the rest of the measures except in the case when the top-1 hashtags are recommended.
    \item Hit rate is always higher than or equal to the scores from other metrics.
    \item The top-$k$ predictions, with $k=1$, is the special case where the hit rate and the hit ratio are the same. When top-$k$ (with $k=1$) prediction.
    \item As can be seen from the results, the F1-score is a very small value close to 0 in the cases where the top-5 and the top-10 hashtags are recommended. This is due to the small value of the precision even if the recall has a bigger value. F1-score penalizes the hits while hit ratio, on the other hand, takes the ratio of the hits ignoring whether the ratio is coming from the recommendation or the ground truth.
\end{itemize}

It is worth mentioning that the Community-Based Hashtag Recommendation model sometimes struggles to recommend any hashtags for some test tweets. The reason is that the model uses a threshold value to find similar tweets and to extract hashtags. However, when no similar tweets can be retrieved, it means that no hashtags can be recommended. Thus, the final score of the model is affected. In the dataset we used, the number of hashtags in the ground truth is not fixed as it ranges from 1 to up to 27 as mentioned in Table~\ref{tab:statistics}. The number of the recommended hashtags is also variable. Hit Ratio is more sensible to this variability in a sense that it computes the ratio of the hits where the hit rate ignores it. As most of the test tweets include a single hashtag, the overall results of the hit ratio and the hit rate are similar especially when the top-1 hashtag is recommended. However, the difference is more evident when the top-5 and the top-10 hashtags are recommended.

For further illustration, Table~\ref{tab:actualexamples} shows six tweets, their ground truth and the top-$3$ recommended hashtags as well as the performance of these recommendations according to the different metrics. The second, fourth, and fifth tweets are examples of the partially correct case. The hit rate scores as 1 even though the hashtags are partially correct. However, their hit ratio varies with the number of hits. In the first, third and sixth examples, the hit rate and hit ratio score the same. All the hashtags in the ground truth are correct in the first and third tweets, and all the hashtags in the recommendation are correct in the sixth example. In the second and third examples, the precision scores are the same even though the numbers of hit are different. Similarly, the recall scores are the same in Examples 2 and 5 with different numbers of hits. The sixth tweet is an example of the fully correct case. The F1-score penalizes the results more than the hit ratio where all three recommendations are correct.

\begin{enumerate}
    \item \textit{Would love an \#iphone app that automatically displays last few tweets/fb updates of anyone i'm about to txt/call}.
    \item \textit{Gorgeous in blue on blue with a dainty swirl this personalized set of \#handmade \#stationary.}
    \item \textit{An interesting question? can real writing happen on a blog? or is "real" only what happens elsewhere? \#realwriter \#writing}
    \item \textit{\#enterprise \#device management isn't simply about opening up the doors to multiple devices and \#operating systems. very one-sided argument.}
    \item \textit{Six quick-hit marketing ideas for \#socialmedia via \#entrepreneur \#smallbiz \#marketing.}
    \item \textit{Abstract original painting \#original \#acrylic \#cmarkandu \#abstract \#blue \#fabulous \#decor.}
    
\end{enumerate}

\begin{table*}[h]
\centering
\resizebox{\textwidth}{!}{%
\begin{tabular}{l|l|l|c|c|c|c|c|}
\cline{2-8}
                        & \multicolumn{1}{c|}{\textbf{Top-3 recommendations}} & \multicolumn{1}{c|}{\textbf{Ground truth}}                                        & \textbf{hit rate} & \textbf{P} & \textbf{R} & \textbf{F1-score} & \textbf{hit ratio} \\ \hline
\multicolumn{1}{|l|}{1} & {[}\textbf{\#iPhone}, \#apple, \#iphoneapp{]}  & {[}\textbf{\#iPhone}{]}                                                             & 1.00                 & 0.33       & 1.00        & 0.50              & 1.00                  \\ \hline
\multicolumn{1}{|l|}{2} & {[}\textbf{\#handmade}, \#epl, \#jewelry{]}                  & {[}\textbf{\#handmade},  \#stationary{]}                                                   & 1.00                 & 0.33       & 0.50        & 0.40              & 0.50                \\ \hline
\multicolumn{1}{|l|}{3} & {[}\textbf{\#realwriter}, \textbf{\#writing}, \#blog{]}     & {[}\textbf{\#realwriter}, \textbf{\#writing}{]}                      & 1.00                 & 0.67       & 1.00        & 0.80              & 1.00               \\ \hline
\multicolumn{1}{|l|}{4} & {[}\textbf{\#enterprise}, \#operating\_systems, \#apple{]}     & {[}\textbf{\#enterprise}, \#device, \#operating{]}                      & 1.00                 & 0.33       & 0.33        & 0.33              & 0.33               \\ \hline
\multicolumn{1}{|l|}{5} & {[}\textbf{\#socialmedia}, \textbf{\#entrepreneur}, \#business{]}     & {[}\textbf{\#socialmedia}, \textbf{\#entrepreneur}, \#smallbiz, \#marketing{]}                      & 1.00                 & 0.67       & 0.50        & 0.57              & 0.67               \\ \hline
\multicolumn{1}{|l|}{6} & {[}\textbf{\#original}, \textbf{\#acrylic}, \textbf{\#decor}{]}                & {[}\textbf{\#original}, \textbf{\#acrylic}, \#cmarkandu, \#abstract, \#blue, \#fabulous, \textbf{\#decor}{]} & 1.00                 & 1.00          & 0.43       & 0.60              & 1.00                  \\ \hline
\end{tabular}
}
\caption{Actual examples of tweets with their top-$3$ recommendations, ground truth, and the scores of the recommendations under various evaluation metrics.}
\label{tab:actualexamples}
\end{table*}

Overall, the hit rate is an inadequate metric. Precision, recall, and F1-score are most affected with different $n_R$ and $n_G$ values even though all the possible matching hashtags are produced. Precision, recall, and F1-score score differently with the same number of correct matching. Hit ratio demonstrates to be a useful metric for its ability to handle the cases of partially correct recommendations consistently.

However, for the case when $n_G < k$, the value of the proposed hit ratio remains unchanged when the value of $k$ increases. This may be considered as undesirable as the metric favours large $k$ values.

\section{Conclusion}
In this paper, we have demonstrated the shortcomings of using hit rate, precision, recall, and F1-score to evaluate hashtag recommendation. We have introduced the hit ratio metric, a new measure that can be used to evaluate hashtag recommendation when each tweet can have a different number of hashtags and when fewer than $k$ hashtags can be generated for top-$k$ recommendation. We have shown that the hit ratio metric is useful. Comparative results of the hit ratio metric and classical evaluation metrics show that the hit ratio, which takes into account the number of ground truth hashtags in each tweet and the actual number of possible hashtags that
can be predicted, is a more sensible metric than the hit rate metric. It also produces more consistent results compared to precision, recall and F1-score. The hit ratio metric can be applied to other problems where the numbers of items in the recommendation and ground truth are variable, for instance, text categorization, product recommendation, and music/movies categorization.

\bibliographystyle{unsrtnat}
\bibliography{references.bib}  

\begin{thebibliography}{22}
\providecommand{\natexlab}[1]{#1}
\providecommand{\url}[1]{\texttt{#1}}
\expandafter\ifx\csname urlstyle\endcsname\relax
  \providecommand{\doi}[1]{doi: #1}\else
  \providecommand{\doi}{doi: \begingroup \urlstyle{rm}\Url}\fi

\bibitem[Mazzia and Juett(2011)]{Mazzia-2011}
Allie Mazzia and James Juett.
\newblock Suggesting hashtags on twitter.
\newblock In \emph{EECS 545 Project, Winter Term, 2011. URL
  http://www-personal.umich.edu/~amazzia/pubs/545-final.pdf}, 2011.

\bibitem[Weston et~al.(2014)Weston, Chopra, and Adams]{Weston-2014}
Jason Weston, Sumit Chopra, and Keith Adams.
\newblock {\#TagSpace:} semantic embeddings from hashtags.
\newblock In Alessandro Moschitti, Bo~Pang, and Walter Daelemans, editors,
  \emph{EMNLP}, pages 1822--1827. ACL, 2014.
\newblock ISBN 978-1-937284-96-1.

\bibitem[Dovgopol and Nohelty(2015)]{Dovgopol-15}
Roman Dovgopol and Matt Nohelty.
\newblock Twitter hash tag recommendation.
\newblock \emph{CoRR}, abs/1502.00094, 2015.

\bibitem[Gong et~al.(2015)Gong, Zhang, and Huang]{Gong-2015-hashtag}
Yeyun Gong, Qi~Zhang, and Xuanjing Huang.
\newblock {H}ashtag {R}ecommendation {U}sing {D}irichlet {P}rocess {M}ixture
  {M}odels {I}ncorporating {T}ypes of {H}ashtags.
\newblock In \emph{Proceedings of the 2015 Conference on EMNLP}, pages
  401--410, Lisbon, Portugal, September 2015. ACL.

\bibitem[Li et~al.(2016{\natexlab{a}})Li, Liu, Jiang, and Zhang]{LiLiu-2016}
Yang Li, Ting Liu, Jing Jiang, and Liang Zhang.
\newblock Hashtag recommendation with topical attention-based lstm.
\newblock In Nicoletta Calzolari, Yuji Matsumoto, and Rashmi Prasad, editors,
  \emph{COLING}, pages 3019--3029. ACL, 2016{\natexlab{a}}.
\newblock ISBN 978-4-87974-702-0.
\newblock URL
  \url{http://dblp.uni-trier.de/db/conf/coling/coling2016.html#LiLJZ16}.

\bibitem[Li et~al.(2016{\natexlab{b}})Li, Xu, He, Deng, and Sun]{LiXu-2016}
Jia Li, Hua Xu, Xingwei He, Junhui Deng, and Xiaomin Sun.
\newblock Tweet modelling with lstm recurrent neural networks for hashtag
  recommendation.
\newblock In \emph{IJCNN}, pages 1570--1577. IEEE, 2016{\natexlab{b}}.
\newblock ISBN 978-1-5090-0620-5.
\newblock URL
  \url{http://dblp.uni-trier.de/db/conf/ijcnn/ijcnn2016.html#LiXHDS16}.

\bibitem[Kywe et~al.(2012)Kywe, Hoang, Lim, and Zhu]{Kywe-2012}
Su~Mon Kywe, Tuan-Anh Hoang, Ee-Peng Lim, and Feida Zhu.
\newblock \emph{{O}n {R}ecommending {H}ashtags in {T}witter {N}etworks}, pages
  337--350.
\newblock Springer Berlin Heidelberg, Berlin, Heidelberg, 2012.
\newblock ISBN 978-3-642-35386-4.
\newblock \doi{10.1007/978-3-642-35386-4_25}.

\bibitem[Wang et~al.(2014)Wang, Qu, Liu, Chen, and Huang]{Wang-2014}
Yuan Wang, Jishi Qu, Jie Liu, Jimeng Chen, and Yalou Huang.
\newblock What to tag your microblog: Hashtag recommendation based on topic
  analysis and collaborative filtering.
\newblock In Lei Chen, Yan Jia, Timos Sellis, and Guanfeng Liu, editors,
  \emph{Web Technologies and Applications}, pages 610--618, Cham, 2014.
  Springer International Publishing.
\newblock ISBN 978-3-319-11116-2.

\bibitem[Sedhai and Sun(2014)]{Sedhai-2014}
Surendra Sedhai and Aixin Sun.
\newblock Hashtag recommendation for hyperlinked tweets.
\newblock In \emph{Proceedings of the 37th International ACM SIGIR Conference
  on Research \&\#38; Development in Information Retrieval}, SIGIR '14, pages
  831--834, New York, NY, USA, 2014. ACM.
\newblock ISBN 978-1-4503-2257-7.
\newblock \doi{10.1145/2600428.2609452}.
\newblock URL \url{http://doi.acm.org/10.1145/2600428.2609452}.

\bibitem[Song et~al.(2015)Song, Meng, and Zheng]{Song-2015}
Shuangyong Song, Yao Meng, and Zhongguang Zheng.
\newblock Recommending hashtags to forthcoming tweets in microblogging.
\newblock In \emph{2015 IEEE International Conference on Systems, Man, and
  Cybernetics}, pages 1998--2003, Oct 2015.
\newblock \doi{10.1109/SMC.2015.348}.

\bibitem[Li et~al.(2016{\natexlab{c}})Li, Xu, He, Deng, and Sun]{Li-2016}
Jia Li, Hua Xu, Xingwei He, Junhui Deng, and Xiaomin Sun.
\newblock Tweet modeling with lstm recurrent neural networks for hashtag
  recommendation.
\newblock In \emph{IJCNN}, pages 1570--1577. IEEE, 2016{\natexlab{c}}.
\newblock ISBN 978-1-5090-0620-5.
\newblock URL \url{http://dblp.uni-trier.de/db/conf/ijcnn/ijcnn2016.html}.

\bibitem[Zhao et~al.(2016)Zhao, Zhu, Jin, and Yang]{Zhao-2016}
Feng Zhao, Yajun Zhu, Hai Jin, and Laurence~T. Yang.
\newblock A personalized hashtag recommendation approach using lda-based topic
  model in microblog environment.
\newblock \emph{Future Gener. Comput. Syst.}, 65\penalty0 (C):\penalty0
  196--206, December 2016.
\newblock ISSN 0167-739X.
\newblock \doi{10.1016/j.future.2015.10.012}.
\newblock URL \url{http://dx.doi.org/10.1016/j.future.2015.10.012}.

\bibitem[Kowald et~al.(2017)Kowald, Pujari, and Lex]{Kowald-2017}
Dominik Kowald, Subhash~Chandra Pujari, and Elisabeth Lex.
\newblock {T}emporal {E}ffects on {H}ashtag {R}euse in {T}witter: {A}
  {C}ognitive-{I}nspired {H}ashtag {R}ecommendation {A}pproach.
\newblock In \emph{Proceedings of the 26th International Conference on WWW},
  pages 1401--1410, Republic and Canton of Geneva, Switzerland, 2017.
  International World Wide Web Conferences Steering Committee.
\newblock ISBN 978-1-4503-4913-0.
\newblock \doi{10.1145/3038912.3052605}.

\bibitem[Alsini et~al.(2017)Alsini, Datta, Li, and Huynh]{Alsini-2017}
Areej Alsini, Amitava Datta, Jianxin Li, and Du~Huynh.
\newblock Empirical analysis of factors influencing twitter hashtag
  recommendation on detected communities.
\newblock In \emph{ADMA - 13th International Conference, {ADMA}, Singapore},
  pages 119--131, 2017.
\newblock \doi{10.1007/978-3-319-69179-4\_9}.
\newblock URL \url{https://doi.org/10.1007/978-3-319-69179-4\_9}.

\bibitem[BenLhachemi and Nfaoui(2018)]{Benlhachemi-2018}
Nada BenLhachemi and El~Habib Nfaoui.
\newblock Using tweets embeddings for hashtag recommendation in twitter.
\newblock \emph{Procedia Computer Science}, 127:\penalty0 7--15, 2018.
\newblock ISSN 1877-0509.
\newblock \doi{https://doi.org/10.1016/j.procs.2018.01.092}.
\newblock URL
  \url{http://www.sciencedirect.com/science/article/pii/S1877050918301030}.
\newblock Proceedings of the First International Conference on ICDS.

\bibitem[Kou et~al.(2018)Kou, Du, Yang, Shi, Cui, Liang, and Geng]{FeiFei-2018}
Fei{-}Fei Kou, Junping Du, Cong{-}Xian Yang, Yan{-}Song Shi, Wan{-}Qiu Cui,
  MeiYu Liang, and Yue Geng.
\newblock Hashtag recommendation based on multi-features of microblogs.
\newblock \emph{J. Comput. Sci. Technol.}, 33\penalty0 (4):\penalty0 711--726,
  2018.
\newblock \doi{10.1007/s11390-018-1851-2}.
\newblock URL \url{https://doi.org/10.1007/s11390-018-1851-2}.

\bibitem[Alsini et~al.(2020)Alsini, Datta, and Huynh]{Alsini-2020}
Areej Alsini, Amitava Datta, and Du~Q. Huynh.
\newblock On utilizing communities detected from social networks in hashtag
  recommendation.
\newblock \emph{IEEE Transactions on Computational Social Systems}, pages
  1--12, 2020.

\bibitem[Sorower(2010)]{Sorower-2010}
Mohammad~S Sorower.
\newblock A literature survey on algorithms for multi-label learning.
\newblock Technical report, Oregon State University, 2010.

\bibitem[Giraldo-Forero et~al.(2015)Giraldo-Forero, Jaramillo-Garzón,
  Castellanos-Domínguez, Ortuño, and Rojas]{Forero-2015}
Andrés~Felipe Giraldo-Forero, Jorge~Alberto Jaramillo-Garzón, César~Germán
  Castellanos-Domínguez, Francisco Ortuño, and Ignacio Rojas.
\newblock Evaluation of example-based measures for multi-label classification
  performance.
\newblock In \emph{Bioinformatics and Biomedical Engineering}, pages 557--564,
  2015.

\bibitem[Li et~al.(2012)Li, Wang, Deng, Wang, and Chang]{UDIdataset-12}
Rui Li, Shengjie Wang, Hongbo Deng, Rui Wang, and {Kevin Chen Chuan} Chang.
\newblock \emph{{T}owards {S}ocial {U}ser {P}rofiling: {U}nified and
  {D}iscriminative {I}nfluence {M}odel for {I}nferring {H}ome {L}ocations},
  pages 1023--1031.
\newblock KDD, 2012.
\newblock ISBN 9781450314626.
\newblock \doi{10.1145/2339530.2339692}.

\bibitem[Mikolov et~al.(2013)Mikolov, Chen, Corrado, and Dean]{Mikolov-2013}
Tomas Mikolov, Kai Chen, Greg Corrado, and Jeffrey Dean.
\newblock Efficient estimation of word representations in vector space.
\newblock \emph{CoRR}, abs/1301.3781, 2013.
\newblock URL
  \url{http://dblp.uni-trier.de/db/journals/corr/corr1301.html#abs-1301-3781}.

\bibitem[Athiwaratkun et~al.(2018)Athiwaratkun, Wilson, and
  Anandkumar]{athiwaratkun-2018}
Ben Athiwaratkun, Andrew Wilson, and Anima Anandkumar.
\newblock Probabilistic {F}ast{T}ext for multi-sense word embeddings.
\newblock In \emph{Proceedings of the 56th Annual Meeting of the ACL}, pages
  1--11, Melbourne, Australia, July 2018. ACL.
\newblock \doi{10.18653/v1/P18-1001}.
\newblock URL \url{https://www.aclweb.org/anthology/P18-1001}.

\end{thebibliography}






\end{document}